\documentclass[useAMS,usenatbib]{mn2e}

\usepackage{graphicx}
\usepackage{wasysym}
\usepackage{amssymb}
\usepackage{stmaryrd}
\usepackage{rotating}
\usepackage{longtable}
\usepackage{lscape}
\usepackage{deluxetable}
\usepackage[usenames,dvipsnames]{xcolor}

\title
[A photometric study of the Open Cluster II: Stellar population and dynamical evolution in NGC\,559]
{A photometric study of the Open Cluster II: Stellar population and dynamical evolution in NGC\,559}

\author[Y.~C.~Joshi et al.]{Y.~C.~Joshi$^{1}$\thanks{E-mail: yogesh@aries.res.in},
L.~A.~Balona$^{2}$,
S.~Joshi$^{1}$,
B.~Kumar$^{1}$,
\\
$^{1}$Aryabhatta Research Institute of Observational Sciences (ARIES), Manora peak, Nainital, India	\\
$^{2}$South African Astronomical Observatory, PO Box 9, Observatory 7935, Cape Town, South Africa	\\
}
\begin{document}

\date{Accepted 07 October 2013 Received 22 July 2013}

\pagerange{\pageref{firstpage}--\pageref{lastpage}} \pubyear{2013}

\maketitle

\label{firstpage}

\begin{abstract}

We present $UBVRI$ photometry of stars in the field of the intermediate-age 
open cluster NGC\,559. By determining the stellar membership probabilities
derived through a photometric and kinematic study of the cluster, we identify
the 22 most probable cluster members. These are used to obtain robust cluster
parameters. The mean proper motion of the cluster is $\mu_x = -3.29\pm0.35$,
$\mu_y = -1.24\pm0.28$ mas yr$^{-1}$. The radial distribution of the stellar
surface density gives a cluster radius of $4'.5\pm0'.2$ (3.2$\pm0.2$ pc).
By fitting solar metallicity stellar isochrones to the colour-colour and 
colour-magnitude diagrams, we find a uniform cluster reddening of 
$E(B-V) = 0.82\pm0.02$.  The cluster has an age of $224\pm25$\,Myr and is 
at a distance of $2.43\pm0.23$\,kpc. From the optical and near-infrared 
two-colour diagrams, we obtain colour excesses in the direction of the 
cluster $E(V-K) = 2.14\pm0.02$, $E(J-K) = 0.37\pm0.01$, and 
$E(B-V)= 0.76\pm0.04$. A total-to-selective extinction of $R_V=3.5\pm0.1$
is found in the direction of the cluster which is marginally higher than the
normal value. We derive the luminosity function and the mass function for the
cluster main sequence. The mass function slope is found to be
$-2.12\pm0.31$. We find evidence of mass segregation in this dynamically
relaxed cluster.

\end{abstract}


\section{INTRODUCTION} \label{sec:intro}
Systematic photometric studies of Galactic open star clusters (OCs) offer unique 
opportunities to understand large-scale star formation processes in the Galaxy 
and in Galactic clusters (Lada 2003).  The precise knowledge of cluster 
parameters such as age, distance, reddening and chemical composition as well as
knowledge of the stellar population distribution and the cluster mass function
at the time of star formation play a key role in understanding the star formation
history. The importance of photometric studies of OCs lies in the colour-colour
and colour-magnitude diagrams derived through multi-band photometric observations. 
Since most of the OCs are embedded in the Galactic disk and are likely to be 
affected by field star contamination, it is essential to discriminate
between members and non-members of the clusters.  The amount of field star 
contamination depends on the location of the cluster.  It is necessary to
perform a detailed membership analysis of the stars found within the observed 
field for a robust investigation of cluster properties (Carraro et al. 2008, 
Yadav et al. 2008). For most of the OCs, kinematical data is unavailable. However,
recent all-sky proper motion catalogues (e.g., Roeser et al. 2010, Zacharias et al.
 2013), provide clues to determine cluster membership.  Together with a photometric
study of the cluster, it becomes possible to draw some conclusions regarding the
dynamical evolution of the cluster.

At ARIES, Nainital, we have been carrying out a long-term observational program 
to search and characterize the variable stars in Galactic open star clusters using
various 1- to 2-m class telescopes in India. The advantage of having such observations
is that they can also be used to study the physical properties of the clusters and
their stellar and dynamical evolution. In Joshi et al. (2012), we performed a
photometric study of the intermediate age open cluster, NGC\,6866, which also
included a search for variable stars in the cluster. The results presented here for
NGC\,559 are a continuation of our efforts to understand star formation in some
unstudied or poorly studied young- and intermediate-age open clusters.

NGC\,559 (RA = 01:29:35, DEC = +63:18:14; $l = 127^\circ.2, b = +0^\circ.75$) is a
moderately populated and heavily reddened intermediate-age 
open cluster, classified as type I$2m$ by Trumpler (1930) and II$2m$ by Ruprecht 
(1966).  It is located in the direction of the second Galactic Quadrant in 
the vicinity of the Perseus and Local arms (Russeil et al. 2007).  Photoelectric
photometry of the cluster was obtained by Lindoff (1969) and Jennens \& Helfer
(1975), while Grubissich (1975) provided photographic photometry of cluster
stars.  A subsequent investigation using CCD photometry was carried out by 
Ann \& Lee (2002, hereafter AL02) and Maciejewski \& Niedzieski (2007, hereafter 
MN07). However, a complete UBVRI study is still lacking.  Moreover, there has not 
been any systematic attempt to identify cluster members in the field of this
cluster.

The main focus of the present study is to accurately determine the fundamental 
parameters of NGC\,559 by identifying cluster members using photometric and 
kinematic criteria.  The outline of the paper is as follows.  A photometric 
study of the cluster is presented in \textsection\,2. The cluster properties are 
discussed in \textsection\,3 and fundamental parameters are derived in 
\textsection\,4. The dynamical study of the cluster is presented in
\textsection\,5. Finally, we discuss the results in \textsection\,6.
\section{Photometric study of the cluster} \label{sec:phot} 
\subsection{Observations and Calibration} \label{sec:photcal}

Johnson-Cousins $UBVRI$ photometry of stars in the field of NGC\,559 was
obtained on 2010, November 30 using the 1-m Sampurnanand telescope at Nainital,
India. The telescope is equipped with a $2k\times2k$ CCD camera which covers a
$\sim 13'\times13'$ field of view. We acquired two frames each in $U$, $B$, $V$,
$R$ and $I$ filters with exposure times of 300, 300, 200, 100, and 60-sec in
respective passbands, respectively, at a typical airmass of about 1.3.  On the
same night we also observed two Landolt's standard fields: SA95 and PG0231+051
(Landolt 1992) at different airmasses. The usual image processing procedures
were performed which included bias subtraction, flat fielding, and cosmic ray
removal.  We used the {\tt IRAF}\footnote{Image Reduction and Analysis Facility
(IRAF) is distributed by the National Optical Astronomy Observatories, which
are operated by the Association of Universities for Research in Astronomy, Inc.,
under cooperative agreement with the National Science Foundation.} software
package for this purpose.
%
%
\begin{figure} 
\includegraphics[width=8.0cm, height=8.0cm]{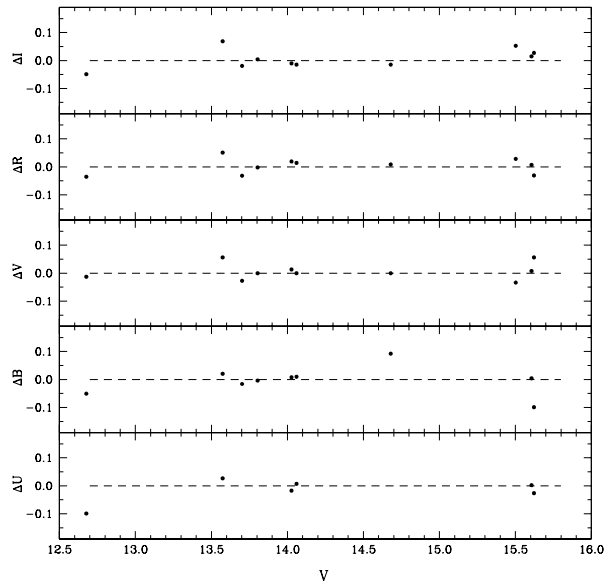} 
\caption{For the standard stars in the Landolt field, plots show residuals of the
differential magnitudes (standard - calibrated) in the $U$, $B$, $V$, $R$, and $I$
bands as a function of $V$ magnitude. The dashed line drawn in each panel represents
a zero difference.}
\label{figure:comp_stand}
\end{figure}
%
\begin{table}
\centering
\caption{Average internal photometric errors per magnitude bin as a function of
brightness.}
\begin{tabular}{llllll}
\hline
mag & $\sigma_U$ & $\sigma_B$   & $\sigma_V$  & $\sigma_R$  & $\sigma_I$\\
\hline
10$-$11 &     -    &    -    &   0.02  &   0.02  &   0.01  \\
11$-$12 &    0.01  &   0.01  &   0.01  &   0.01  &   0.01  \\
12$-$13 &    0.02  &   0.01  &   0.01  &   0.01  &   0.01  \\
13$-$14 &    0.01  &   0.00  &   0.01  &   0.01  &   0.01  \\
14$-$15 &    0.01  &   0.01  &   0.01  &   0.01  &   0.01  \\
15$-$16 &    0.01  &   0.01  &   0.01  &   0.01  &   0.02  \\
16$-$17 &    0.02  &   0.01  &   0.01  &   0.01  &   0.02  \\
17$-$18 &    0.04  &   0.01  &   0.02  &   0.02  &   0.04  \\
18$-$19 &    0.08  &   0.02  &   0.03  &   0.03  &   0.07  \\
19$-$20 &    0.20  &   0.04  &   0.06  &   0.06  &   0.20  \\
20$-$21 &     -    &   0.07  &   0.11  &   0.13  &    -    \\
21$-$22 &     -    &   0.21  &   0.26  &    -    &    -    \\

\hline
\end{tabular}
\end{table}

Photometry of the frames was performed using the {\tt DAOPHOT II} profile fitting
software (Stetson 1987).  Details of the photometric calibration obtained on this
night are given in Joshi et al. (2012). Transformation coefficients for the standard
stars were determined as follows. \\
\\
   $ u = U + 8.16\pm0.01 - (0.05\pm0.01)(U-B) + (0.55\pm0.02)X $ \\
   $ b = B + 5.81\pm0.02 - (0.01\pm0.02)(B-V) + (0.29\pm0.03)X $\\
   $ v = V + 5.43\pm0.01 - (0.08\pm0.01)(B-V) + (0.15\pm0.01)X $\\
   $ r = R + 5.23\pm0.01 - (0.09\pm0.02)(V-R) + (0.09\pm0.02)X $\\
   $ i = I + 5.63\pm0.02 + (0.01\pm0.01)(R-I) + (0.07\pm0.02)X $\\

\noindent
where $u, b, v, r$ and $i$ are the aperture instrumental magnitudes and $U$, $B$,
$V$, $R$ and $I$ are the standard magnitudes and $X$ is the airmass. The
difference between the calibrated magnitudes derived from the above transformation
equations and the Landolt (1992) magnitudes are plotted in Fig.~1.
The standard deviations of these measurements are estimated to be 0.04, 0.05, 0.03,
0.03, and 0.03\,mag for the $U$, $B$, $V$, $R$ and $I$ filters, respectively. The
above transformation coefficients were used to convert instrumental magnitudes to the
standard system. The average internal photometric error per magnitude bin in all the
five filters on the night of standardization are listed in Table\,1. This shows that
photometric errors become large ($>0.1$\,mag) for stars fainter than $V \approx 20$ mag.
To standardize the data on remaining nights, differential photometry was performed
using a linear fit between the standard and instrumental magnitudes on each night,
assuming that most of the stars are non-variable.
%
\begin{table*}
\centering
\caption{Completeness Factor of the photometric data in $U$, $B, V, R$ and $I$ bands in cluster and field regions.}
\begin{tabular}{c|cc|cc|cc|cc|cc}
\hline
mag &\multicolumn{2}{c|}{$U$}&\multicolumn{2}{c|}{$B$}&\multicolumn{2}{c|}{$V$}&\multicolumn{2}{c|}{$R$}&\multicolumn{2}{c|}{$I$} \\
range &cluster&field&cluster&field&cluster&field&cluster&field&cluster&field\\
\hline
10$-$11  &    -  &    -   &    -   &    -   &   -    &   -    &  1.00  &  1.00  &  0.83  &  0.83\\
11$-$12  & 1.00  & 1.00   & 1.00   &  1.00  &  0.94  & 1.00   &  1.00  &  1.00  &  1.00  &  1.00\\
12$-$13  & 1.00  & 0.96   & 1.00   &  1.00  &  1.00  & 1.00   &  1.00  &  0.95  &  0.97  &  1.00\\
13$-$14  & 1.00  & 1.00   & 1.00   &  1.00  &  1.00  & 0.96   &  1.00  &  1.00  &  0.95  &  1.00\\
14$-$15  & 0.95  & 1.00   & 1.00   &  0.96  &  1.00  & 1.00   &  0.90  &  1.00  &  1.00  &  0.97\\
15$-$16  & 0.97  & 1.00   & 1.00   &  1.00  &  0.97  & 1.00   &  0.97  &  0.96  &  0.97  &  0.97\\
16$-$17  & 0.97  & 0.97   & 0.97   &  1.00  &  0.93  & 1.00   &  0.97  &  0.97  &  0.89  &  0.92\\
17$-$18  & 1.00  & 0.97   & 0.93   &  1.00  &  1.00  & 0.97   &  0.90  &  0.97  &  0.89  &  0.88\\
18$-$19  & 0.65  & 0.70   & 1.00   &  0.94  &  0.93  & 0.97   &  0.82  &  0.83  &  0.46  &  0.51\\
19$-$20  & 0.19  & 0.20   & 0.90   &  0.94  &  0.86  & 0.91   &  0.72  &  0.73  &   -    &   -\\
20$-$21  &  -    &  -     & 0.45   &  0.52  &  0.48  & 0.52   &  0.09  &  0.16  &   -    &   -\\
\hline
\end{tabular}
\end{table*}

%
\subsection{Completeness of the data}
It is necessary to determine the completeness of the data as it is not always possible 
to detect every star in the CCD frame, particularly the faintest stars.  The
completeness factor (CF) is required in order to derive the luminosity function and 
the mass function of the cluster as well as to estimate the stellar density
distribution. The {\tt ADDSTAR} routine in {\tt DAOPHOT} was used to determine CF.
This involves adding randomly selected artificial stars with
different, but known, magnitudes and positions to the original frames. We
added about 10--15\% of the actually detected stars, so that the crowding 
characteristics of the original image is almost unchanged. We added simulated stars 
to all bands in such a way that they have similar geometric locations.  We
varied the brightness of the artificial star depending on its location
relative to the Main-Sequence (MS) in the $V$ band. We constructed five frames
for each passband and re-processed them with the same 
procedure as used in the original frames. The average ratio of number of stars 
recovered to the number of simulated stars in the different magnitude bins
gives the  CF as a function of magnitude.  The CF in all five passbands for both 
cluster and field regions is given in Table\,2.  From the table, one can see
that the completeness decreases towards the fainter stars because of the
increased crowding caused by the large number of low-mass stars.
\subsection{Astrometry} \label{sec:photast}
In order to transform CCD pixel coordinates to celestial coordinates, we used 
the on-line digitized ESO catalogue included in the {\tt skycat} software as an 
absolute astrometric reference frame.  A linear astrometric solution was derived 
for the $V$ filter reference frame by matching positions of 63 well
isolated, bright stars in the USNOA2.0 catalogue. The {\tt ccmap} and {\tt cctran} 
routines in {\tt IRAF} were used to find a transformation equation which
gives the celestial coordinates $(\alpha, \delta)$ as a function of the pixel 
coordinates, $(X,Y)$.  The resulting celestial coordinates have standard
deviations of 0.1\,arcsec in both right ascension and declination. 

A finding chart for stars in NGC\,559 is shown in Fig.~2.
We do not see any significant concentration of stars at the center which suggests
that cluster is loosely bound.
%
\begin{figure} 
\includegraphics[scale=0.4]{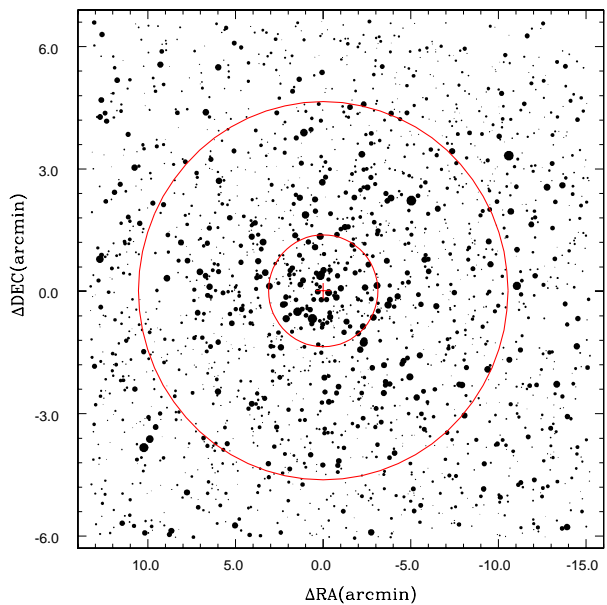} 
\caption{Finding chart of stars in the field of NGC\,559. North is upwards and 
East is on the left. The sizes of the filled circles are proportional to the
brightness of the stars in the $V$ band.  The faintest are $V = 21$.  The
inner and outer rings indicate core and cluster radii with origin, $(0,0)$, 
at the cluster center.}
\label{figure:fchart} 
\end{figure}
%
\subsection{Comparison with previous photometry} \label{sec:photcomp} 
Photoelectric and photographic observations of NGC\,559 have been carried out by 
Lindoff (1969) and Grubissich (1975) respectively.  Photographic magnitudes
contain relatively large errors, while photoelectric magnitudes are mostly confined
to stars brighter than $V \sim 15$, hence we did not compare them with our
photometry in the present study. CCD photometry in the $UBVRI$ bands is
discussed in AL02, but these data have not been published.  Recently, MN07 performed 
a wide field CCD survey of a few clusters using a 90/180\,cm Schmidt-Cassegrain 
telescope equipped with a SBIG camera. This survey also includes NGC\,559, for which 
$BV$ data are presented, but only for stars brighter than about 18\,mag.  

%
\begin{figure} 
\includegraphics[scale=0.4]{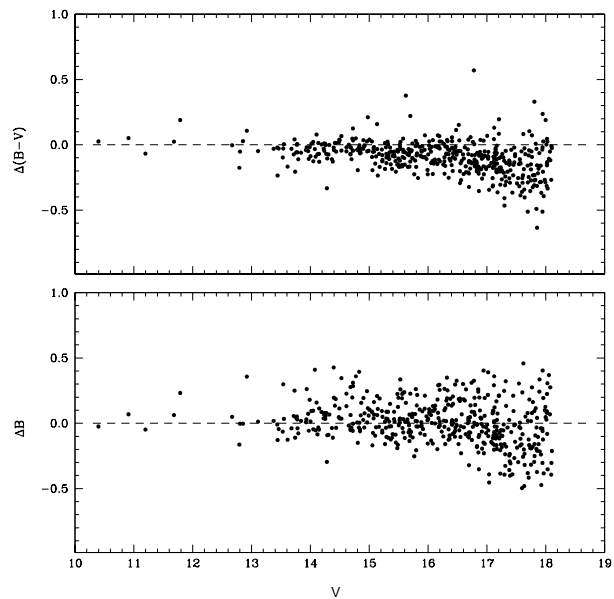} 
\caption{Differences, $\Delta$ between measurements presented in MN07 and in
the present study for $B$ magnitude and $(B-V)$ colour.  Zero difference is indicated by the dashed line.}
\label{figure:comp_phot} 
\end{figure}
%
\begin{table}
\centering
\caption{Differences in $B$ and $(B-V)$ between MN07 and present study. The standard deviation in the difference for each magnitude bin is also given in the bracket.}
\begin{tabular}{ccc}
\hline
V mag range &  $\Delta$B       & $\Delta$(B-V) \\
\hline
10$-$11     &    0.02 (0.05)   &    0.04 (0.01) \\
11$-$12     &    0.08 (0.12)   &    0.05 (0.11) \\
12$-$13     &    0.05 (0.17)   &   -0.02 (0.09) \\
13$-$14     &    0.03 (0.11)   &   -0.06 (0.06) \\
14$-$15     &    0.07 (0.14)   &   -0.03 (0.07) \\
15$-$16     &    0.02 (0.12)   &   -0.07 (0.09) \\
16$-$17     &    0.03 (0.15)   &   -0.09 (0.11) \\
17$-$18     &   -0.08 (0.20)   &   -0.17 (0.15) \\
18$-$19     &   -0.09 (0.24)   &   -0.21 (0.20) \\
\hline
\end{tabular}
\end{table}

%
We found 1112 stars in the MN07 catalogue which are included in our study.  However,
there are only 687 stars in common for which both $B$ and $V$ magnitudes are available. 
We have cross-identified stars in the two catalogues on the assumption that stars
are correctly matched if the difference in position is less than $1\arcsec$.  On this 
basis, we found 505 stars in common which have similar $B$ and $V$ magnitudes
within 0.5\,mag.  A comparison of $B$ magnitudes and $(B-V)$ colours between 
the two catalogues is shown in Fig.~3. The mean difference and 
standard deviation in each magnitude bin is given in Table\,3. This shows that our $B$ 
measurements are in fair agreement with those given in the MN07 catalogue. However, 
there is a systematic difference in $(B-V)$ colours between the two catalogues.
\subsection{A complete $UBVRIJHK$-proper motion catalog}
We have compiled a photometric catalogue of 2393 stars in the field of NGC\,559. 
The catalogue contains 515, 1288, 2177, 2352 and 2221 stars measured in the
$UBVRI$ bands respectively. Near-infrared magnitudes for point sources around 
NGC\,559 have also been obtained from the Two Micron All-sky survey (2MASS; 
Skrutskie et al. 2006). The 2MASS provides photometry in the $J$ (1.25 $\mu$m), 
$H$ (1.65 $\mu$m) and $K_s$ (2.17 $\mu$m) bands up to a limiting magnitude of 15.8, 15.1, and 14.3 
respectively. We found $JHK_s$ magnitudes for 917 stars in the field of
NGC\,559, of which 906 stars are identified in our catalogue within $1\arcsec$
of their positions.  The $K_s$ magnitudes were converted into 
$K$ magnitudes using equations given in the Carpenter et al. (2001). The proper 
motions have been taken from Roeser et al. (2010) which gives a
catalogue for about 900 million stars derived from the USNOB1.0 and 2MASS all sky catalogues.
%
\begin{sidewaystable}
\caption{:Photometric catalogue of 2293 stars detected in the field of cluster NGC~559. The error in magnitudes indicates the internal photometric error in the measurement. Table is sorted in the order of increasing $V$ magnitude. Column 1 gives identification number and columns 2 and 3 give right ascension and declination for J(2000). From columns 4 to 11, we provide photometric magnitudes and corresponding error in the $UBVRIJHK$ passbands. Columns 12 and 13 gives proper motion in x- and y- directions.}
\label{tab:photclus}
\centering
\tiny
\begin{tabular}{ccccccccccccccrr}
\hline
  ID  &    RA       &     DEC      &         U        &        B         &        V         &       R          &        I         &        J         &         H        &         K        &      $\mu_x$      &    $\mu_y$     \\
\hline
     1& 01:29:12.19 &  +63:20:28.1 & 10.952$\pm$0.002 & 10.838$\pm$0.004 & 10.397$\pm$0.032 & 10.188$\pm$0.011 &  9.838$\pm$0.004 &  9.595$\pm$0.021 &  9.446$\pm$0.026 &  9.418$\pm$0.022 & -11.5$\pm$1.9  &   3.4$\pm$1.9\\
     2& 01:28:49.96 &  +63:21:34.2 & 11.295$\pm$0.011 & 11.016$\pm$0.009 & 10.459$\pm$0.027 &        -         &  9.729$\pm$0.026 &  9.378$\pm$0.023 &  9.181$\pm$0.028 &  9.117$\pm$0.020 &   9.9$\pm$1.9  &   4.6$\pm$1.9\\
     3& 01:29:34.77 &  +63:17:34.3 & 11.448$\pm$0.011 & 11.484$\pm$0.004 & 10.908$\pm$0.012 & 10.552$\pm$0.014 & 10.233$\pm$0.006 &  9.856$\pm$0.021 &  9.576$\pm$0.024 &  9.550$\pm$0.024 &  10.5$\pm$1.9  &   3.0$\pm$1.9\\
     4& 01:30:13.25 &  +63:14:24.7 & 14.340$\pm$0.012 & 12.819$\pm$0.004 & 11.196$\pm$0.014 & 10.257$\pm$0.020 &  9.369$\pm$0.019 &  8.139$\pm$0.017 &  7.358$\pm$0.039 &  7.210$\pm$0.005 &   7.8$\pm$2.7  &  -7.8$\pm$2.7\\
     5& 01:29:38.19 &  +63:17:44.6 & 12.814$\pm$0.010 & 12.539$\pm$0.002 & 11.680$\pm$0.008 & 11.198$\pm$0.009 & 10.774$\pm$0.006 &        -         &        -         &        -         &  27.5$\pm$2.7  &  -1.3$\pm$2.7\\
     6& 01:29:32.98 &  +63:18:37.5 & 12.195$\pm$0.019 & 12.377$\pm$0.022 & 11.787$\pm$0.008 & 11.431$\pm$0.009 & 11.029$\pm$0.012 &        -         &        -         &        -         & -12.3$\pm$4.5  &  -4.0$\pm$4.6\\
     7& 01:28:48.12 &  +63:18:22.6 & 12.165$\pm$0.021 &        -         & 11.928$\pm$0.006 &        -         & 11.146$\pm$0.020 &        -         &        -         &        -         &  -4.9$\pm$3.1  &  -6.7$\pm$3.1\\
     8& 01:29:23.07 &  +63:16:58.7 & 17.131$\pm$0.027 & 14.780$\pm$0.006 & 12.382$\pm$0.007 & 10.207$\pm$0.017 &        -         &  5.797$\pm$0.015 &  4.707$\pm$0.015 &  4.264$\pm$0.013 &  -4.6$\pm$14.1  & -76.0$\pm$14.1\\
     9& 01:30:11.87 &  +63:14:37.2 & 13.552$\pm$0.012 & 13.424$\pm$0.007 & 12.670$\pm$0.009 & 12.214$\pm$0.011 & 11.759$\pm$0.021 &        -         &        -         &        -         &  20.9$\pm$4.9  &  -4.3$\pm$5.1\\
    10& 01:29:36.72 &  +63:22:08.0 & 17.523$\pm$0.038 & 14.927$\pm$0.011 & 12.792$\pm$0.009 & 11.457$\pm$0.020 & 10.065$\pm$0.005 &  8.460$\pm$0.021 &  7.518$\pm$0.037 &  7.215$\pm$0.027 &   0.8$\pm$5.0  &  -1.3$\pm$5.0\\
    11& 01:30:23.27 &  +63:19:02.0 & 16.061$\pm$0.012 & 14.477$\pm$0.010 & 12.804$\pm$0.011 & 11.796$\pm$0.009 & 10.777$\pm$0.016 &  9.496$\pm$0.019 &  8.771$\pm$0.028 &  8.575$\pm$0.020 &  10.8$\pm$19.6  & -65.4$\pm$19.6\\
    12& 01:29:36.36 &  +63:20:06.9 & 13.941$\pm$0.005 & 13.571$\pm$0.003 & 12.853$\pm$0.008 & 12.445$\pm$0.011 & 11.951$\pm$0.007 & 11.369$\pm$0.017 & 11.136$\pm$0.024 & 11.057$\pm$0.022 &  -1.3$\pm$2.7  &   8.0$\pm$2.7\\
    13& 01:29:17.49 &  +63:17:54.5 &        -         & 14.537$\pm$0.016 & 12.920$\pm$0.010 & 11.964$\pm$0.009 & 10.977$\pm$0.007 &  9.702$\pm$0.021 &  9.002$\pm$0.026 &  8.792$\pm$0.022 &  15.4$\pm$25.7  & -51.8$\pm$25.7\\
    14& 01:29:40.77 &  +63:17:34.6 & 16.577$\pm$0.015 & 14.922$\pm$0.004 & 13.109$\pm$0.009 & 12.031$\pm$0.010 & 10.935$\pm$0.007 &  9.561$\pm$0.019 &  8.730$\pm$0.026 &  8.510$\pm$0.022 &  -8.6$\pm$14.1  & -23.7$\pm$14.1\\
    15& 01:29:12.54 &  +63:16:08.6 & 15.026$\pm$0.013 & 14.467$\pm$0.004 & 13.375$\pm$0.006 & 12.720$\pm$0.010 & 11.951$\pm$0.007 & 11.063$\pm$0.019 & 10.663$\pm$0.026 & 10.519$\pm$0.024 &  -5.5$\pm$4.0  &  -0.6$\pm$4.0\\
    16& 01:29:20.05 &  +63:18:23.1 & 14.511$\pm$0.013 & 14.150$\pm$0.008 & 13.431$\pm$0.011 & 12.999$\pm$0.011 & 12.471$\pm$0.014 & 11.807$\pm$0.021 & 11.628$\pm$0.028 & 11.492$\pm$0.026 & -12.0$\pm$4.0  &  -1.5$\pm$4.0\\
    17& 01:28:39.03 &  +63:15:58.0 & 15.158$\pm$0.010 & 14.455$\pm$0.011 & 13.437$\pm$0.014 & 12.862$\pm$0.016 & 12.198$\pm$0.014 &        -         &        -         &        -         &  34.0$\pm$4.0  &   0.8$\pm$4.0\\
    18& 01:29:31.07 &  +63:18:12.5 & 16.387$\pm$0.016 & 15.061$\pm$0.004 & 13.442$\pm$0.009 & 12.491$\pm$0.011 & 11.514$\pm$0.007 & 10.282$\pm$0.019 &  9.583$\pm$0.024 &  9.400$\pm$0.022 & -10.8$\pm$6.4  &  -2.2$\pm$6.4\\
    19& 01:29:32.99 &  +63:19:35.4 & 14.719$\pm$0.008 & 14.293$\pm$0.004 & 13.460$\pm$0.008 & 12.960$\pm$0.010 & 12.394$\pm$0.010 & 11.707$\pm$0.019 & 11.456$\pm$0.024 & 11.353$\pm$0.024 &  -2.2$\pm$4.0  &  -5.1$\pm$4.0\\
    20& 01:29:46.02 &  +63:19:27.0 & 17.038$\pm$0.026 & 15.350$\pm$0.005 & 13.529$\pm$0.007 & 12.427$\pm$0.007 & 11.306$\pm$0.009 &  9.877$\pm$0.019 &  9.076$\pm$0.024 &  8.861$\pm$0.022 &  -2.3$\pm$5.0  &   0.0$\pm$5.0\\
      &             &              &                  &                  &                  &                  &                  &                  &                  &                  &                 &               \\
      &             &              &                  &                  &                  &                  &                  &                  &                  &                  &                 &               \\
  2392& 01:29:14.55 &  +63:20:49.8 &        -         &        -         &        -         & 19.632$\pm$0.101 & 18.148$\pm$0.064 &        -         &        -         &        -         &        -        &        -      \\
  2393& 01:29:11.31 &  +63:18:23.3 &        -         &        -         &        -         & 19.324$\pm$0.073 & 18.267$\pm$0.048 &        -         &        -         &        -         &        -        &        -      \\
\hline
\end{tabular} 
\end{sidewaystable}

The $UBVRIJHK$ magnitudes and proper motions, wherever measured, are presented in
Table\,4, sorted in increasing order of $V$. In the catalogue, column 1 contains
a running number, columns 2 and 3 give right ascension and declination for J(2000),
columns 4 to 11 provide photometric magnitudes and corresponding errors in the $UBVRIJHK$
passbands. The proper motion along the RA and DEC directions and their respective errors 
are given in the columns 12 and 13. Only a short extract of Table\,4 is
shown; the complete catalogue is available at the WEBDA open cluster data base 
website\footnote{http://obswww.unige.ch/webda/} or can be obtained directly from 
the authors.
\section{Structural Properties of the cluster}\label{pcl}
\subsection{Spatial Structure: Radial density profile} \label{rdp}
The spatial structure and precise center of the star cluster is difficult to 
determine due to the irregular shape of the cluster and the non-uniform 
distribution of stars at different brightness levels.  We define the cluster
centre as the region where maximum stellar density is attained.  To determine 
this value, we consider all stars with $V < 19$ for which the completeness level
is in excess of 90\%. We found that the stellar density peaks at the pixel
coordinate (510, 535), corresponding to a cluster centre at ($\alpha, \delta$)
= (01:29:32.33, +63:18:14.5). An error of up to $10\arcsec$ is expected in
locating the cluster center.

To draw the radial density profile (RDP), we determined the stellar density in 
concentric rings, $0'.5$ wide, centered on the cluster center. The errorbars 
were derived assuming Poisson statistics. We fitted a King (1966) stellar density 
as modified by Kaluzny \& Udalski (1992):
\begin{center}
$\rho(r) = \rho_f + \frac{\rho_0}{1+ (\frac{r}{r_c})^2} $
\end{center}
\noindent
Here $\rho_f$ is the field density and $r_c$ is the core radius of cluster where 
the stellar density, $\rho(r)$, becomes half of its central value, $\rho_0$. The 
stellar density distribution in the $V$ band is shown in Fig.~4. 
A $\chi^2$ best fit to the radial density profile is shown in the figure along
with the field star density. The cluster boundary is considered to be the point
in the radial direction when $\rho(r)$ falls below the field star density by
3$\sigma$.  The value of the core radius was found to 
be $1'.3\pm0'.3$ and the cluster radius was estimated to be $4'.5\pm0'.2$. Our 
radius estimate is the same as that determined by AL02. The inner and outer rings 
in Fig.~2 represent the core and cluster regions, respectively.
%
\begin{figure} 
\includegraphics[scale=0.4]{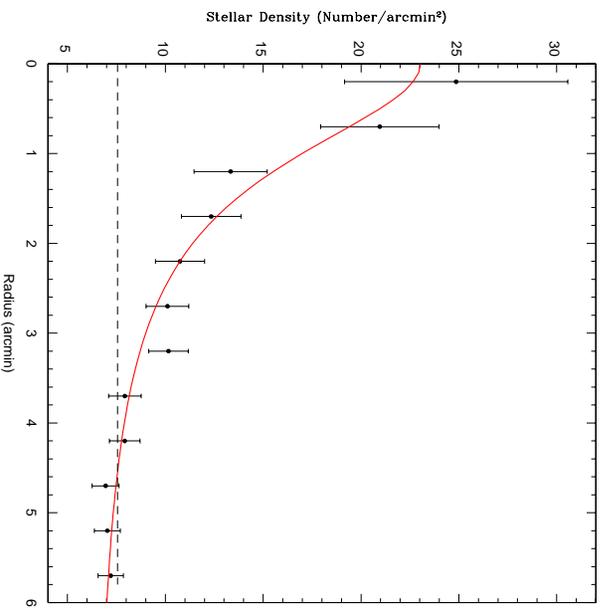} 
\caption{The stellar density distribution in NGC\,559 for stars brighter than
19\,mag. The solid line represents the King profile while the horizontal dashed 
line indicates the field density.}
\label{figure:rdp} 
\end{figure}
%

We noticed that the core radii derived from bright stars is smaller than
those which include stars up to $V$=20.  This suggests that: i) the core and 
cluster radii derived using the RDP are only approximate or, ii) that
there is mass segregation due to the dynamical evolution of the cluster.
In the latter case, it would seem that bright massive stars move towards the
cluster centre, while faint low-mass stars move away from the cluster center.  A 
similar trend has been noticed by Lee et al. (2013) in their investigation of the 
clusters NGC\,1245 and NGC\,2506. A detailed study on the dynamical evolution
is presented in Section 5.
\subsection{Colour-Magnitude Diagram} \label{cmd}
%
\begin{figure*} 
\includegraphics[height=12cm, width=15cm]{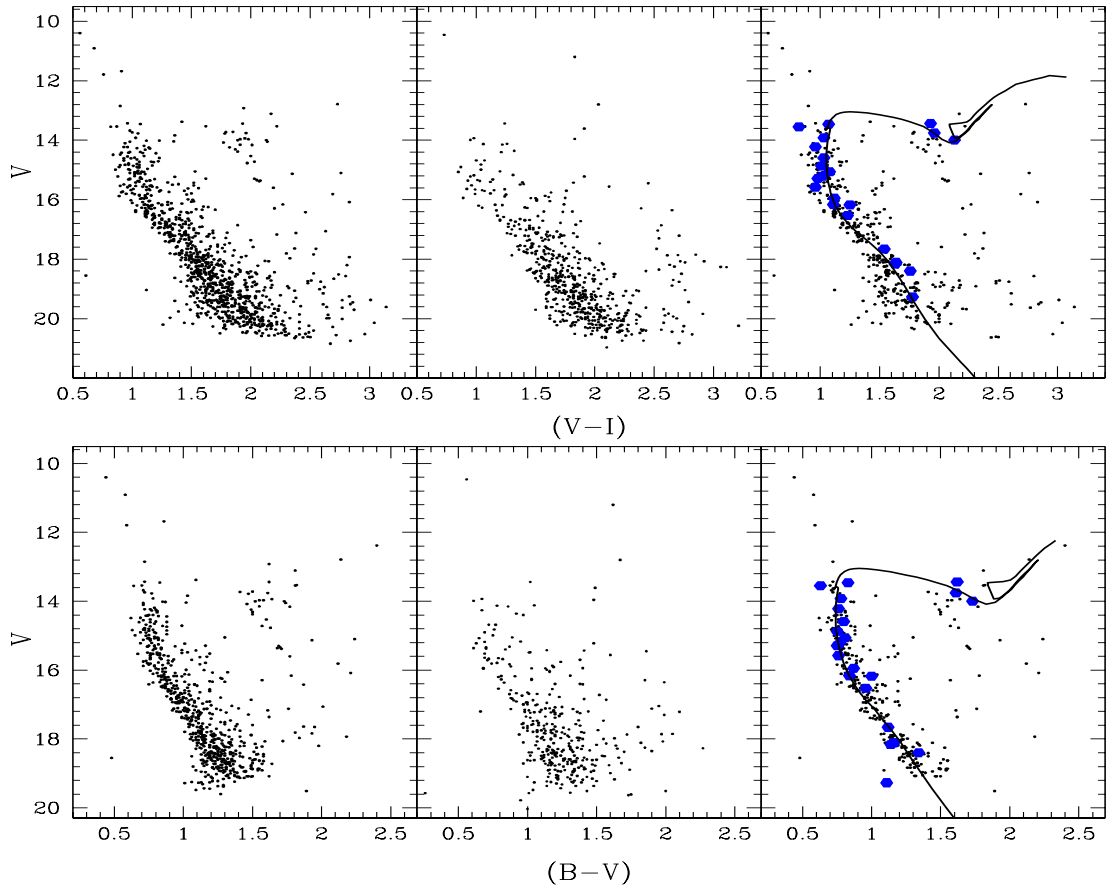} 
\caption{The $(B-V)/V$ and $(V-I)/V$ CMDs for stars in the cluster region (left panels) and field region (middle panels). The right panels show the same for the 
stars in the cluster field after statistical subtraction of field stars.
The most probable cluster members (see, Table\,5) are shown by large filled circles. The solid line represents the best fit isochrone to the cluster MS for $\log$(Age)=8.35 (see, \textsection\,\ref{cmd} for detail).}
\label{figure:cmd} 
\end{figure*}

The identification of the cluster main sequence in the colour-magnitude
diagrams (CMDs) allows a model-dependent mass, radius, and distance for each star
to be determined. To draw the CMD, we used the area within cluster radius ($4'.5$) as the
{\it `cluster region'} and an equal area outside the cluster radius of $5'.6$
as the {\it `field region'}. In the left panels of Fig.~5,
we constructed calibrated $(B-V)$, and $(V-I)$ vs $V$ diagrams of NGC~559
using the stars falling in the cluster region. A similar diagram for
the stars in the field region are shown in the middle panels of the same figure.

Since stars in the cluster region are contaminated by the field
star population, we adopted a statistical approach to remove the field star
contamination.  This method is based on a comparison of the cluster 
and field CMDs.  We removed all cluster stars in the $(V-I)$/$V$ CMD which
fall within a grid cell of $(V, V-I)$ = ($\pm 0.25$, $\pm 0.125$) of the field
stars CMD. A similar removal process was done for the $(B-V)$/$V$ CMD with
a grid of $(V, B-V)$ = ($\pm 0.25$, $\pm 0.10$). We iterated the procedure for 
each star lying on the CMDs of the field region. We were finally left with 462 
stars in the $(V-I)$/$V$ CMD and 341 stars in the $(B-V)$/$V$ CMD. We found more 
stars in the $(V-I)$/$V$ CMD because our photometry goes deeper in the $V$ and $I$ 
bands than in the $B$ band. The statistically cleaned cluster CMDs are shown in the 
right hand panels of Fig.~5. The
spatial distribution of stars extracted after the statistical subtraction shows that
the inner region is dominated by giant and upper-MS stars, whereas the outer region is 
dominated by low-mass stars. The lack of stars in some pockets is quite evident
in the cleaned CMDs. These kind of gaps in MS are not unusual and have been found
in many clusters (see detail in Rachford \& Canterna 2000). AL02 also noticed a
gap at $M_V\sim3.5$ mag ($m_v\sim18.1$) in the cluster NGC\,559 similar to
the one seen in the present study. This suggests that these gaps could be due to a real lack
of cluster members in some magnitude bins.
\subsection{Mean proper motion} \label{pm}
Recently, Roeser et al. (2010) provided a catalogue which lists stellar 
coordinates with an accuracy of 80--300\,mas and absolute proper motion 
with an accuracy of 4--10\,mas yr$^{-1}$ for about 900 million stars. A 
cross-match of these stars with our catalogue using a matching
criterion of $1\arcsec$ resulted in 1824 stars in common. In Table\,4, we
provide proper motions of these stars along the RA and DEC directions and their
respective errors. Fig.~6 shows the proper motion distribution in the RA-DEC plane. 

To determine the mean proper motion of the cluster, we considered those 341 stars
which fall in both the cleaned $(V-I)$/$V$ and $(B-V)$/$V$ CMDs. Among them,
307 stars were found  within $1\arcsec$ of the Roeser et al. (2010) catalogue
positions.  We determined the mean and $\sigma$ values of the proper motion in 
both RA and DEC directions and rejected those stars which fall outside 3$\sigma$ 
in both the directions. We iterated this procedure until all values fall within 
3$\sigma$ of the mean. We were finally left with 229 stars which were used to 
determine the mean proper motion of the cluster NGC~559. These stars are shown 
by filled circles in Fig.~6. The mean proper motion of the cluster 
determined in this way is\\

\noindent $\bar{\mu}_{x} = -3.29\pm0.35$\,mas\,yr$^{-1}$; ~~~~
$\bar{\mu}_{y} = -1.24\pm0.28$\,mas\,yr$^{-1}$\\

%
\begin{figure} 
\includegraphics[scale=0.4]{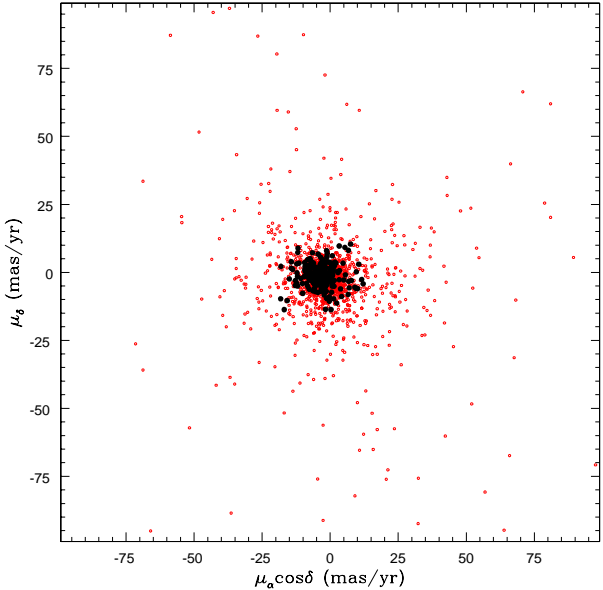} 
\caption{The distribution of stars in the $\mu_{x}$ - $\mu_{y}$ plane for which proper motion values are determined in our catalogue and given in Table\,4. The 229 stars used to estimate the mean proper motion are shown by filled circles.}
\label{figure:pm} 
\end{figure}
\noindent where the uncertainties are standard deviations. A similar matching criteria using UCAC4 catalogue (Zacharias et al. 2013) has given only 167 stars though UCAC4 catalogue provides proper motion with higher accuracy. A 3$\sigma$ clipping analysis done on the proper motions left 145 stars which resulted a mean proper motion of $\bar{\mu}_{x} = -4.45\pm0.49$ and $\bar{\mu}_{y} = 1.65\pm0.37$\,mas\,yr$^{-1}$ in RA and DEC directions, respectively. The proper motion for the cluster NGC\,559 estimated using two different catalogues are therefore in close agreement within their given uncertainties. From the radial-velocity measurements of 24 stars computed from the data of the Tycho-2 catalog, Loktin \& Beshenov (2003) estimated a proper motion of $\bar{\mu}_{x} = -1.59\pm0.41$ and $\bar{\mu}_{y} = -0.52\pm0.46$\,mas\,yr$^{-1}$ for the cluster NGC~559, which is lower than the present estimates.
\subsection{Probable cluster members} \label{pk}
Open clusters are mostly located within the densely populated Galactic plane and 
often contaminated with large numbers of field stars belonging to the disc population.
It is therefore essential to discriminate between members and non-members in order to obtain
correct cluster parameters. To identify the most-likely cluster members in NGC~559,
we first derive different membership probabilities for each star in the cluster
field based on their spatial distribution, position in the colour-magnitude diagram
and proper motions.
\subsubsection{Spatial probability} \label{sp}
The spatial probability, $P_{\rm sp}$, is a function of the angular distance of 
the star from the cluster centre, $r$, and is given by
$$P_{\rm sp} = 1-\frac{r}{r_c}$$
where $r_c$ is the angular radius of the cluster. Using $r_c = 4'.5$ derived in
\textsection\,\ref{rdp}, we  determined $P_{\rm sp}$ for all the 960 stars falling within 
the cluster radius.  For $r \geq r_c$ we assign $P_{\rm sp} = 0$.  We found 176 stars 
within the core region of the cluster for which $P_{\rm sp} \geq 0.67$.
\subsubsection{Statistical probability} \label{sp}
We determine statistical probability which is based on a comparison of the cluster 
CMD with that of the field CMD, as discussed in \textsection\,\ref{cmd}. In this 
method we removed all the stars in the $(B-V)$/$V$ CMD of the cluster field which 
fall within a grid cell of $(V, B-V)$ = ($\pm 0.25$, $\pm 0.10$), in the field CMD.
After iterating the procedure for each star lying on the CMD of the field region,
we found 341 stars for which we assigned statistical probabilities $P_{\rm st}=1$.
For the remaining stars, we assigned $P_{\rm st}=0$.
\subsubsection{Kinematic probability} \label{pk}
The kinematic probability, $P_{k}$, is defined as the deviation in the proper 
motion of stars in both RA and DEC directions with respect to the mean proper
motion of the cluster.

Using the method given by Kharchenko et al. (2004), we determined $P_k$ for each 
star using
$$ P_k = \exp \left\{-0.25  \left[ (\mu_{x} - \bar{\mu}_{x})^{2}/ \sigma_{x}^{2} +
(\mu_{y} - \bar{\mu}_{y})^{2}/ \sigma_{{y}}^{2} \right]  \right\}$$
where $\sigma_{x}^{2} = \sigma_{\mu_{x}}^{2}+\sigma_{\bar{\mu}_{x}}^{2}$ and 
$\sigma_{y}^{2} = \sigma_{\mu_{y}}^{2}+\sigma_{\bar{\mu}_{y}}^{2}$.
The mean proper motion of the cluster NGC\,559 is taken from our analysis carried out
in \textsection\,\ref{pm}. We found 1824 stars for which $P_k$ could be
estimated using the Roeser et al. (2010) catalogue. 


\begin{table}
\caption{The list of 22 most-likely members identified in the cluster NGC\,559. These
stars were identified using various probability criteria (see, \textsection\,\ref{pk}
for detail.)}
\label{tab:clusmem}
\centering
\begin{tabular}{ccccc|ccc}
\hline
   RA       &    DEC       &   V     &  (B-V) & (V-I)  \\
   (J2000)  &    (J2000)   &   (mag) &  (mag) & (mag)  \\
\hline
01:29:31.07 & +63:18:12.5  &  13.44  &  1.62  &  1.93  \\
01:29:32.99 & +63:19:35.4  &  13.46  &  0.83  &  1.07  \\
01:29:44.60 & +63:18:22.2  &  13.55  &  0.63  &  0.82  \\
01:29:20.89 & +63:17:36.4  &  13.76  &  1.61  &  1.96  \\
01:29:22.77 & +63:18:20.2  &  13.92  &  0.78  &  1.03  \\
01:29:34.17 & +63:19:19.6  &  14.00  &  1.73  &  2.13  \\
01:29:28.30 & +63:18:19.0  &  14.22  &  0.76  &  0.96  \\
01:29:36.93 & +63:18:30.6  &  14.59  &  0.80  &  1.03  \\
01:29:29.19 & +63:19:37.8  &  14.87  &  0.75  &  1.01  \\
01:29:32.73 & +63:18:02.3  &  15.06  &  0.81  &  1.08  \\
01:29:25.03 & +63:18:41.6  &  15.21  &  0.77  &  1.02  \\
01:29:29.05 & +63:18:37.6  &  15.29  &  0.75  &  0.98  \\
01:29:21.37 & +63:18:03.1  &  15.58  &  0.76  &  0.96  \\
01:29:38.52 & +63:18:02.2  &  15.95  &  0.87  &  1.12  \\
01:29:33.45 & +63:16:56.4  &  16.16  &  0.84  &  1.11  \\
01:29:27.26 & +63:18:47.5  &  16.18  &  1.00  &  1.25  \\
01:29:39.44 & +63:17:15.2  &  16.53  &  0.96  &  1.23  \\
01:29:40.51 & +63:18:54.0  &  17.66  &  1.12  &  1.54  \\
01:29:34.39 & +63:17:18.7  &  18.11  &  1.16  &  1.64  \\
01:29:32.42 & +63:16:53.7  &  18.16  &  1.14  &  1.63  \\
01:29:37.57 & +63:18:44.9  &  18.40  &  1.34  &  1.76  \\
01:29:29.26 & +63:19:01.4  &  19.27  &  1.11  &  1.78  \\
\hline                                      	
\end{tabular}
\end{table}

To identify the most-likely members in the cluster NGC~559, we considered
stars those lie in the core region of the cluster ($P_{\rm sp} \geq 0.67$), fall
within the cleaned CMD ($P_{\rm st}$=1.0), and proper motion within
1$\sigma$ of the mean proper motion ($P_{\rm k} \geq 0.60$). We identified 22 such
stars in our catalogue which fulfill above criteria. These criteria are conservative
in the sense that they confer membership status to the selected stars, but it does
not mean that other stars are non-members. The positions of these stars along with
their magnitude, and colours are given in Table\,5. To determine robust cluster 
parameters for NGC\,559, these stars were preferentially used in our analysis as 
explained in the following section.
\section{Cluster Parameters} \label{cp}
\subsection{Reddening law and two-colour-diagrams}\label{eltcd}
Though the normal reddening law, $R_V = \frac{A_V}{E(B-V)} = 3.1$, is valid 
for lines of sight that do not pass through dense clouds (Sneden et al. 1978), 
clusters associated with gas and dust or behind the dusty Galactic spiral arms 
may give a different value of $R_V$. To investigate the nature of the reddening 
law, Chini \& Wargau (1990) showed that the TCDs of the form $(\lambda-V)/(B-V)$ 
can be used, where $\lambda$ is any broad-band filter. The slope of the TCD 
distinguishes normal extinction produced by grains in the diffuse interstellar 
medium from that caused by abnormal dust grains (Pandey et al. 2000). We studied 
the reddening law in the cluster NGC\,559 by drawing $(\lambda-V)/(B-V)$ 
diagrams for the $\lambda = R$, $I$, $J$, $H$ and $K$ bands as shown in
Fig.~7. The slope, $m_{\rm cluster}$, was determined by fitting
a linear relation in the TCD for the stars in the cluster region and a best fit
determined after a 3$\sigma$-clipping iteration. The estimated values of $m_{\rm cluster}$ 
for all five colours are given in Table\,6 along with their normal values. To 
derive the value of total-to-selective extinction $R_{\rm cluster}$ in the 
direction of NGC\,559, we used the approximate relation (cf. Neckel \& Chini 1981)
\begin{center}
$R_{\rm cluster} = \frac{m_{\rm cluster}}{m_{\rm normal}} \times R_{\rm normal}$
\end{center}
\noindent Using $R_{\rm normal} = 3.1$, we estimated $R_{\rm cluster}$ in 
different passbands to be $3.1 < R_{\rm cluster} < 3.5$ which is marginally 
higher than the normal value. The reddening law in the direction of the cluster 
is found to be normal at longer wavelengths but anomalous at shorter wavelengths.  
%
\begin{figure}  
\includegraphics[width=8.8cm, height=15cm]{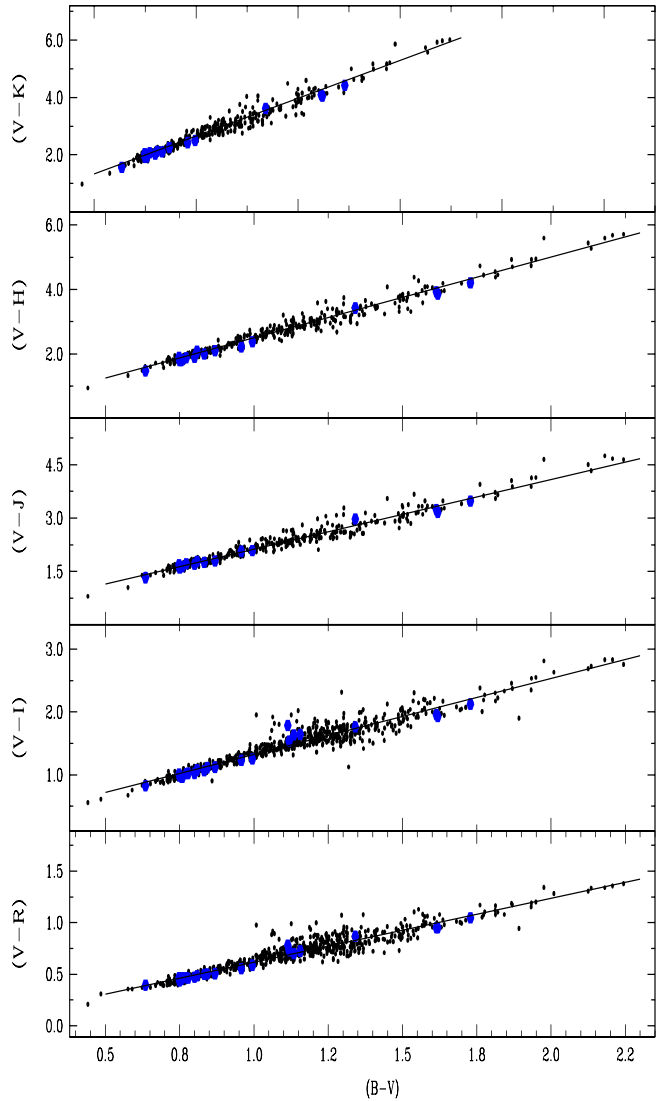} 
\caption{The $(\lambda-V)/(B-V)$ two-colour diagram for the stars within cluster
region. The most probable cluster members are shown by large filled circles. The
continuous lines represent the slope determined through least square linear fit.}
\label{figure:tcd} 
\end{figure}
%
\begin{table}
\caption{The slopes of the $(\lambda-V)/(B-V)$ diagrams in the direction of the cluster NGC~559. Normal value in the same colour is given in the bracket.}
\label{tcd}
\begin{tabular}{ccccc}
\hline \noalign{\smallskip}
$\frac{(R-V)}{(B-V)}$ & $\frac{(I-V)}{(B-V)}$ & $\frac{(J-V)}{(B-V)}$ & $\frac{(H-V)}{(B-V)}$& $\frac{(K-V)}{(B-V)}$  \\
\hline
0.62$\pm$0.01&1.21$\pm$0.01 &1.95$\pm$0.02&2.50$\pm$0.02&2.63$\pm$0.02 \\
(0.55)       &   (1.10)     &     (1.96)  &    (2.42)   &   (2.60)    \\
\hline
\end{tabular}
\end{table}

%
%
\subsection{Reddening determination: $(U-B)$ vs $(B-V)$ TCD} \label{ccd}
The reddening, $E(B-V)$, in the cluster region is normally determined using the 
$(U-B)$/$(B-V)$ two-colour diagram (TCD). Out of 2393 stars in our catalogue, we found 
only 501 stars for which all the $U$, $B$ and $V$ magnitudes are available. Among them, 
we considered only 275 stars within the cluster which have a $U$ band photometric error 
less than 0.05. The resulting TCD is shown in Fig.~8. As mentioned in
the previous section, the normal reddening law is not applicable at shorter wavelengths.
Therefore, we have fitted intrinsic zero-age main sequence (ZAMS) isochrones of
solar metallicity (Marigo et al. 2008) to the observed MS stars by shifting $E(B-V)$
and $E(U-B)$ along different values of the reddening vector $\frac {E(U-B)} {E(B-V)}$.
A visual inspection shows that the best fit is achieved for
$\frac {E(U-B)} {E(B-V)} = 0.84\pm0.01$. This gives a mean reddening of
$E(B-V) = 0.82\pm0.02$ in the direction of NGC\,559 as shown by a solid line in
Fig.~8. In determining the reddening, we used only
stars having colours corresponding to spectral classes earlier than A0 because stars
having later spectral types are more affected by metallicity and background
contamination (Hoyle et al. 2003). The colour excess obtained in the present study
is in good agreement with the value $E(B-V) = 0.81\pm0.05$ given by AL02, but higher
than $0.68^{+0.11}_{-0.12}$ obtained by MN07. Using the Johnson \& Morgan (1953)
$Q$-method for stars earlier than A0 ($(B-V)<0.84$), we determined the reddening of
each star. The reddening distributions of these stars show that reddening is uniform
over the whole cluster.

Considering $R_{\rm normal} = 3.1$, we estimated a higher value of $R_{\rm cluster}=3.6$
for ultraviolet wavelengths. This further suggests an anomalous reddening law at shorter
wavelengths in the direction of NGC\,559. Chini \& Wargau (1990) pointed out that both
larger and smaller size grains may increase $R_{\rm cluster}$. However, some of the
recent studies (e.g., Whittet et al. 2001, Pandey et al. 2008 and references therein)
suggest that a value of $R_{\rm cluster}$ higher than the normal is indicative of the
presence of larger dust grains. As NGC\,559 is situated behind the Perseus arm, a high
reddening and anomalous reddening law is not surprising.
%
\begin{figure}
\includegraphics[scale=0.4]{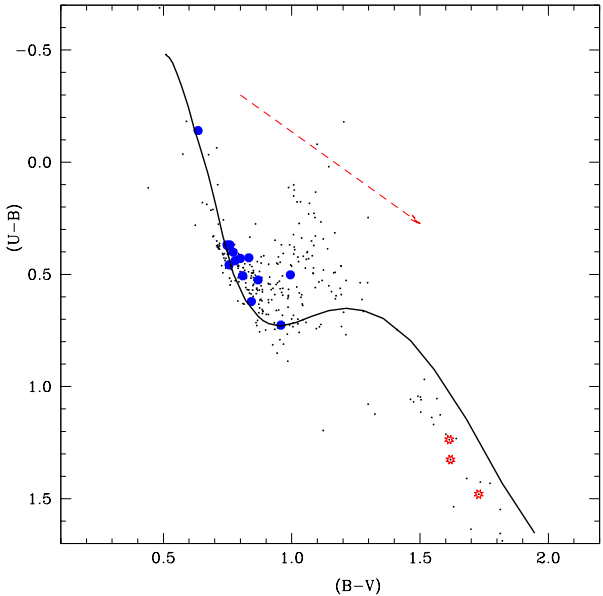} 
\caption{The $(U-B)$ versus $(B-V)$ diagram for the stars in NGC\,559. The small 
dots represent the stars which lie within the cluster boundary and have a
$U$-band photometric error less than 0.05. Filled circles are the most
probable cluster members. The 3 stars shown in red colour represent red giants belonging to the cluster but not considered in the reddening estimation. The thick dashed arrow represents the slope (0.84) and direction of the reddening vector. The solid line represents the ZAMS with solar metallicity taken from the Marigo et al. (2008) shifted for $E(B-V)=0.82$.}
\label{figure:ccd} 
\end{figure}
%
\subsection{Distance and Age determination} \label{cmd}
The distance and age of NGC\,559 can be estimated by visual fitting of theoretical 
isochrones to the MS.  For this purpose we used $(B-V)/V$ and $(V-I)/V$ CMDs shown
in the right panels of Fig.~5. The stars show a broad but clearly 
distinct MS in the CMD. The width is mainly caused by cluster binaries and 
field stars. There are a few stars scattered towards 
the red side of the CMDs. We suspect these may be foreground field stars
which have remained due to incomplete subtraction of the field star 
contamination.  We presume most of them belong to the Perseus spiral arm. In order 
to obtain the most reliable estimates of the cluster parameters, we identified those 
stars in the cleaned CMDs which lie inside the core region and have proper motions 
within 1$\sigma$ of the mean proper motion of the cluster. These stars are shown by
the blue filled circles in Fig.~5. We used stellar evolutionary 
isochrones published by the Padova group\footnote{http://pleiadi.pd.astro.it/} 
(Marigo et al. 2008) to estimate the cluster age and distance. We fixed
the reddening to the value estimated in \textsection\,\ref{ccd}. A simultaneous 
best fit was made of the isochrones in the bluest envelope of the $(B-V)/V$ and 
$(V-I)/V$ CMDs, corrected for a mean reddening of $E(B-V)=0.82$ and $E(V-I)=1.12$
assuming $\frac{E(V-I)}{E(B-V)}$=1.37 (Schlegel et al. 1998). This gives an age 
of $\log$(Age)=$8.35\pm0.05$ and an apparent distance modulus of $(m-M)$ 
= $14.80\pm0.05$ for NGC\,559. The errors in age and distance are strongly 
influenced by a few blue and red supergiants in the CMDs.

As we have seen in \textsection\,\ref{eltcd} and \textsection\,\ref{ccd},
that the total-to-selective extinction in optical region varies from 3.4 to 3.6. 
We adopted a mean value of $R_V = 3.5\pm0.1$ as the total-to-selective extinction 
in the direction of NGC\,559. Assuming a total extinction of $A_V = R_V \times E(B-V)$, 
the reddening-free distance modulus is estimated as $(V_0 - M_V)$ = $11.93\pm0.20$, 
which corresponds to a distance of $2.43\pm0.23$\,kpc for NGC\,559. The linear 
diameter of the cluster is estimated to be $6.4\pm0.4$\,pc. Since the cluster lies 
very close to the Galactic plane, a large foreground extinction of about $E(B-V)=0.56$ 
is expected in that direction (Schlegel  et al. 1998, Joshi 2005). 

The position of NGC\,559 in Galactic coordinates is $l = 127^\circ.2, b = +0^\circ.75$.
Assuming that the Sun is at a distance of 8.5\,kpc from the Galactic center, 
the Galactocentric rectangular coordinates of NGC\,559 are $X\sim1.88$\,kpc, 
$Y\sim1.44$\,kpc, $Z\sim+30.9$\,pc and a Galactocentric distance of $\sim$
10.1\,kpc for the cluster. This places NGC\,559 just outside the Perseus spiral 
arm. The distance of the cluster from the Galactic plane is 
smaller than the typical scale height of the thin disk ($\approx$ 75 pc). 
This is in agreement with Joshi (2007) which found that most of the OCs younger
than about 300\,Myr lie somewhere within $\pm$100\,pc of the Galactic Plane. 
%
\begin{figure}
\includegraphics[scale=0.4]{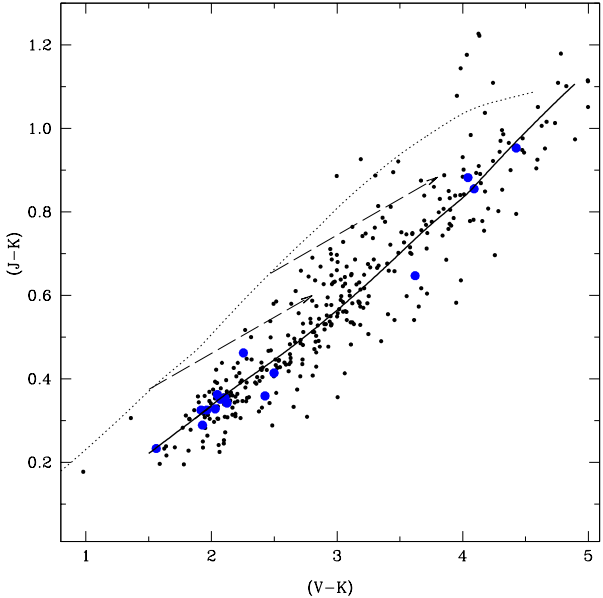}
\caption{The $(V-K)/(J-K)$ colour-colour diagram for stars within the cluster
boundary. Filled circles are the most
probable cluster members. The dotted line is the solar metallicity isochrone for
$\log$(Age)=8.35, while the two dashed lines indicate the direction of the normal 
reddening vector. The solid line is obtained by using reddenings of
$E(V-K) = 2.14$ and $E(J-K) = 0.37$.}
\label{figure:irtcd} 
\end{figure}    
%
\subsection{Interstellar extinction in the near-infrared} \label{IRebv}
To determine interstellar extinction in the near-IR, we used 370 stars for 
which $VJK$ magnitudes were available in our catalogue.  The $(V-K)/(J-K)$ diagram 
is shown in Fig.~9. We used the normal reddening law for the 
infra-red colours, as given in Table\,6, and shifted the stars along the 
reddening vector $\frac{E(J-K)}{E(V-K)} = 0.173$ using solar metallicity
isochrones given by the Marigo et al. (2008). The best fit to points in the
$(V-K)/(J-K)$ diagram gives a colour excess of 
$E(V-K) = 2.14\pm0.02$ and $E(J-K) = 0.37\pm0.01$ by minimizing $\chi^2$. The 
theoretical isochrone shifted by the above values is shown by the solid line in 
Fig.~9. Using the Whittet \& van Breda (1980) relation for 
$R_K = 1.1 E(V-K)/E(B-V)$, which is insensitive to the reddening law, we
obtained $E(B-V) = 0.76\pm0.04$ for the reddening in NGC\,559. This is close 
to $E(B-V) = 0.82\pm0.02$ determined using the $(U-B)/(B-V)$ TCD. The agreement 
between two complementary methods suggests that our values are robust.

The fundamental parameters derived for NGC\,559 in this study are summarized in 
Table\,7. 
%

\begin{table}
\caption{Fundamental parameters of the cluster NGC 559 as determined in the present study.}
\label{tab:funpar}
\begin{tabular} {ll}
\hline
Cluster parameters      &      \\ \hline
RA(2000) & 01:29:32.33 \\
DEC(2000) & +63:18:14.5 \\
$R_{core}$ & $1.3\pm0.3$ arcmin  \\
$R_{cluster}$ & $4.54\pm0.20$ arcmin\\
Mean $E(B-V)$      & 0.82$\pm$0.02 mag \\
$(m-M)_v$          & $14.8\pm0.05$ mag    \\
$\bar{\mu}_{\alpha}$ & -3.29$\pm$0.35 ~mas/yr \\
$\bar{\mu}_{\delta}$ & -1.24$\pm$0.28 ~mas/yr \\
log(Age/yr)        & 8.35$\pm$0.05       \\ 
Total-to-selective extinction $R_V$ & $3.5\pm0.1$ \\
Distance           & $2.43\pm0.23$ kpc  \\  
Diameter           & $6.4\pm0.4$ pc \\
\hline										      
\end{tabular}									    
\end{table}									    

%
\subsection{Comparison to previous results}
NGC\,559 has been studied in the past by various authors.  Lindoff (1969) found 
it to be a very old cluster with an age of about 1000\,Myr, while Jennens \& Helfer 
(1975) estimate the age at only 100\,Myr. Both studies used photoelectic
photometry. Grubissich (1975), Lynga (1987), AL02 and MN07, all estimated
the cluster age at $\log$(Age)$ =8.7\pm0.1$. In this paper we used only
the most probable cluster members to estimate $\log$(Age)=$8.35\pm0.05$.

The distance of the cluster is estimated to be about 1.3\,kpc (Lindoff 1969), 
6.3\,kpc (Jennens \& Helfer 1975), and 1.15\,kpc (Lynga 1987). The recent CCD study 
by AL02 and MN07 determined a distance of $2.3\pm0.3$ and
$2.17^{+0.56}_{-0.82}$\,kpc respectively.  The latter value is close to the
distance of $2.43\pm0.22$ kpc determined in the present study.  Previous estimates 
of reddening, $E(B-V)$, are about 0.45 (Lindoff 1969), 0.62$\pm$0.17 (Jennes \& 
Halfer 1975), 0.54 (Lynga 1987), and $0.68^{+0.11}_{-0.12}$ (MN07). However, AL02 
obtained a higher value of $E(B-V)=0.81\pm0.05$, which is in good agreement with 
our value of $0.82\pm0.02$.
\section{Dynamical Study of the cluster} \label{ds}
The dynamical properties of the cluster can be studied by determining the
luminosity and mass functions of the cluster members.

\subsection{Luminosity function} \label{lf}
The luminosity function (LF) is the total number of cluster members in different 
magnitude bins. After correcting for the data completeness to both cluster and field regions, the 
number of probable cluster members was obtained by subtracting the contribution 
of field stars from stars in the cluster region. The estimated number of stars 
in each magnitude bin for both the cluster $(N_C)$ and field regions $(N_F)$ are 
given in Table\,8. To determine the photometric LFs in $(V-I)/V$ and $(B-V)/V$ CMDs,
we subtracted $N_F$ from $N_C$ and resultant probable members $(N_P)$ are
given in the 4th and 7th columns of Table\,8, respectively.
\subsection{Mass function} \label{mf}
The initial mass function (IMF) is defined as the distribution of stellar masses 
per unit volume in a star formation event. Along with the star formation rate, 
the IMF determines the subsequent evolution of clusters (Kroupa 2002). Since the 
direct determination of the IMF is not possible due to the dynamical evolution of 
stellar systems, we derive the mass function (MF), which is the relative number of 
stars per unit mass and can be expressed by a power law $N(\log M) \propto M^{\Gamma}$. 
The slope, $\Gamma$, of the MF can be determined from
$$\Gamma = \frac{d\log N(\log\it{m})}{d\log\it{m}}$$
\begin{table}		
\centering		
\caption{Luminosity Functions of the stars in the $(V-I)/V$  and $(B-V)/V$ CMDs of the cluster and field regions. $N_C$ and $N_F$ denote the number of stars in a magnitude bin in the cluster and field regions. $N_P$ ($N_C - N_F$) gives the difference between MS stars between the cluster and field regions.}
\begin{tabular}{c|ccc|c|cc|ccc|c }		
\hline		
$V$ range  & \multicolumn{3}{|c|}{(V-I)/V CMD} & & & \multicolumn{3}{|c|}{(B-V)/V CMD}  \\		
  (mag)    &$N_C$&$N_F$&$N_P$&&    &$N_C$&$N_F$&$N_P$   \\		
\hline		
12$-$13&     0 &   0  &   0  & &    &	3   &	0   &	3    \\
13$-$14&     8 &   2  &   6  & &    &	5   &	0   &	5    \\
14$-$15&     8 &   0  &   8  & &    &  12   &	2   &  10    \\
15$-$16&    24 &   6  &  18  & &    &  36   &	5   &  31    \\
16$-$17&    68 &  14  &  54  & &    &  66   &  11   &  55    \\
17$-$18&    73 &  33  &  40  & &    &  48   &  33   &  15    \\
18$-$19&    91 &  41  &  50  & &    &  65   &  30   &  35    \\
19$-$20&    80 &  38  &  42  & &    &	9   &	4   &	5    \\
20$-$21&    46 &  38  &   8  & &    &	-   &	-   &	-    \\
\hline		
\end{tabular}		
\end{table}		

%
%
\begin{figure} 
\includegraphics[height=14cm, width=9cm]{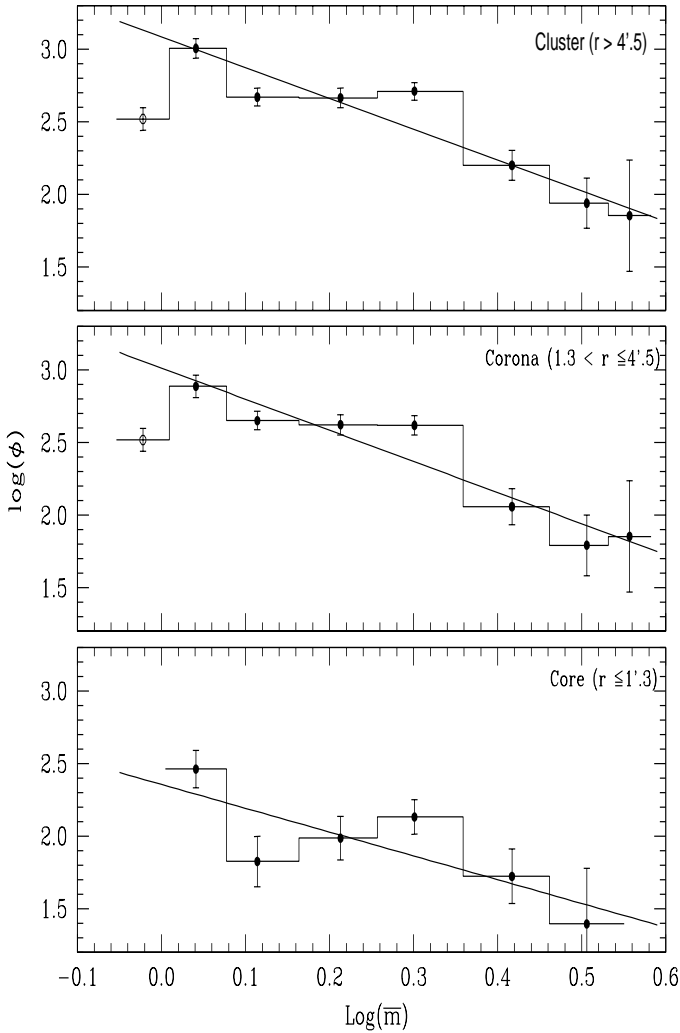} 
\caption{MF derived for the core region (lower panel), corona (middle panel), and
whole cluster region (upper panel). The error bars represent $1/\sqrt N$ errors. The
continuous line is the fit to the data excluding the points shown by open circles.}
\label{figure:mf} 
\end{figure}

\noindent  where $N\log(m)$ is the number of stars per unit logarithmic mass.
The masses of probable cluster members can be determined by comparing
observed magnitudes with those predicted by a stellar evolutionary model
if the age, reddening, distance and metallicity are known. 
%
\begin{table}
\centering
\caption{The mass function of the cluster NGC 559 derived from $(V-I)/V$ CMD.}
\begin{tabular}{cccccc}
\hline
$V$ Range&Mass Range&$\overline{m}$&$N$&$log(m)$& $log(\phi)$ \\
(mag)    &($M_\odot$)  &($M_\odot$)&     &          &                 \\
\hline
13-14  & 3.67-3.44  &   3.61  &   2   &  0.557  &   1.854 \\
14-15  & 3.44-2.86  &   3.21  &   7   &  0.506  &   1.939 \\
15-16  & 2.86-2.20  &   2.61  &  18   &  0.417  &   2.200 \\
16-17  & 2.20-1.73  &   2.00  &  53   &  0.301  &   2.710 \\
17-18  & 1.73-1.40  &   1.63  &  43   &  0.213  &   2.665 \\
18-19  & 1.40-1.10  &   1.30  &  49   &  0.114  &   2.670 \\
19-20  & 1.10-1.00  &   1.10  &  42   &  0.041  &   3.006 \\
20-21  & 1.00-0.80  &   0.95  &  32   & -0.022  &   2.519 \\
\hline
\end{tabular}
\end{table}

As seen in Fig.~5, the $(V-I)/V$ CMD goes deeper than the
$(B-V)/V$ CMD, so we used the former to determine the MF of the cluster. 
The main factors that limit the accuracy of the MF are incompleteness and 
field star contamination.  While the central region of the cluster may be affected 
by data incompleteness, the outer region is more likely to be affected by 
field star contamination. After statistically correcting for the field star
contamination, we determined the MF in three regions i.e., 
the core region ($r\leq1'.3$), the corona ($1'.3 < r \leq 4'.5$), and the whole 
cluster region ($r\leq4'.5$). The MF determined for the cluster region is given 
in Table\,9. Fig.~10 shows the MF in the cluster fitted for the 
MS stars with masses $0.8 \leq M/M_\odot < 3.7$. The error bars were calculated
assuming Poisson statistics.  In determining the slope, we have considered
only those data points which are shown by filled circles in Fig.~10.
The slope of the MF ($\Gamma$) in the mass range $1.0 \leq M/M_\odot < 3.7$ in
each region is calculated using a least square method and shown by the solid
line in the figure. Table\,10 summarizes the MF slopes in the cluster for all
three regions. 
%
\begin{table}
\centering
\caption{MF slope ($\Gamma$) for the stars in the mass range of about 1.0 to 3.7 $M_\odot$}
\begin{tabular}{l|c}
\hline
Region &   $\Gamma$ \\
(Radial distance)  &  \\
\hline
core ($r\leq1'.3$)        & -1.64$\pm$0.62  \\
corona ($1'.3<r\leq4'.5$) & -2.14$\pm$0.30  \\
cluster ($r\leq4'.5$)     & -2.12$\pm$0.31  \\
\hline
\end{tabular}
\end{table}

For the mass range $0.4 < M/M_\odot < 10$, the classical value derived by
Salpeter (1955) for the MF slope is -1.35.  The MF slope in the core region 
is in agreement with the Salpeter MF slope within the given uncertainty, but
it is steeper for the corona and cluster regions.
This suggests a preferential distribution of relatively massive stars towards 
the central region of the cluster. When we determined the MF slopes for two extreme
age limits of the cluster considering the uncertainty in our age determinations,
we found that the MF slope is slightly dependent on the age of the cluster, and
varies by a maximum of $\sim$20\% .

It is worth pointing out that the mass range for probable MS stars in this cluster
is quite small.  It is possible that some of the low mass stars may have escaped
from the cluster as a result of stellar encounters 
between stars of different masses. On the other hand, the initially massive stellar 
members of the cluster have now evolved and may possibly be white dwarfs or have
undergone supernova explosions.  Very deep photometry will be required to
detect white dwarfs or supernova remnants, if present.
\subsection{Mass Segregation} \label{ms}
There is ample proof of mass segregation in star clusters, i.e. a tendency of 
higher-mass stars to approach towards inner region and lower-mass stars towards
outer region of the cluster. This appears to be a result of equipartition of
energy through stellar encounters (e.g., Mathieu \& Latha, 1986, Sagar et al. 1988, 
Pandey et al. 2001).
To understand if mass segregation is an imprint of the star formation process in 
the cluster and/or a result of dynamical evolution, we determined the dynamical
relaxation time, $T_E$.  This is the time in which individual stars in the cluster 
exchange energies and their velocity distribution approaches the Maxwellian 
equilibrium. It can be expressed as
\begin{center}
$T_E = \frac{8.9 \times 10^5 (N R_h^3/\bar{m})^{1/2}} {\log(0.4N)}$\\
\end{center}
\noindent where $T_E$ is in Myr, $N$ is the total number of cluster members, 
$R_h$ is the radius (in parsecs) containing half of the cluster mass and 
$\bar{m}$ is mean mass of the cluster members in solar units (cf. Spitzer 
\& Hart, 1971). We estimated a total of 202 MS stars in the mass range 
$0.8 \le M/M_\odot < 3.7$. The total mass of the cluster is obtained by subtracting 
the total stellar mass in the field region from the cluster region.  This
results in a total mass of $\sim 344 M_\odot$ for NGC\,559, which gives an 
average mass of $\sim 1.7 M_\odot$ per star.  The contribution of the low-mass 
stellar population is critical for constraining the total cluster mass, which is 
crucial in understanding the dynamical evolution and the long-term survival 
of a cluster (e.g., de Grijs \& Parmentier 2007, and references therein). 
We can not rule out the possibility of poor subtraction of field 
stars from the cluster or an observing bias against detecting low-mass stars
which might result in underestimating the total mass of the cluster. Therefore 
the present value may be taken as a lower limit for the cluster mass, while 
the estimated mean stellar mass can be taken as an upper limit.

It can be seen that the half-radius of the cluster, $R_h$, plays an important 
role in the determination of the dynamical relaxation time, $T_E$. 
Unfortunately, this quantity is unknown for most clusters and is generally taken 
as half of the total cluster radius. Nevertheless, we can estimate $R_h$
by taking advantage of the statistical removal of field stars 
from the field region and knowledge of the approximate stellar masses using
stellar isochrones. The value of $R_h$ determined in this way is $\sim
2.3$\,pc, which is $\sim 70\%$ of the cluster radius. A $R_h$ value larger
than half of the cluster radius suggests that inner region has a
deficiency of massive stars which have now evolved. We estimated the
dynamical relaxation 
time $T_E$ = 19.2\,Myr for NGC\,559. However, cluster members fainter than the 
limiting $V$ magnitude of our observations results in a decrease of $N$ and 
an increase of $\bar{m}$, leading to an underestimation of $T_E$. Therefore, 
$T_E$ obtained in this way should be regarded as a lower limit. The values used 
in the estimation of $T_E$ are summarized in Table\,11. $T_E$ determined in the 
present study is much smaller than the present age of about 224\,Myr. We 
conclude, therefore, that NGC\,559 is a dynamically relaxed cluster.
%
\begin{table}
\centering
\caption{Parameters used to estimate $T_E$ for the cluster NGC 559.}
\label{tab:dyna}
\begin{tabular}{ll}
\hline
Probable members ($N$) & 202           \\
Total cluster mass & 344 $M_\odot$     \\
Cluster half radius ($R_h$) & 2.3 pc   \\
Mean stellar mass ($\bar{m}$) & 1.7 $M_\odot$ \\
Dynamical relaxation time ($T_E$) & 19.2 Myrs     \\
$Age$ of the cluster & 224 Myrs      \\
\hline
\end{tabular}
\end{table}

%
\section{Conclusion and Summary}
We present results of an ongoing photometric survey in order to determine
the structure, and astrophysical and dynamic evolution parameters of the intermediate
age galactic cluster NGC\,559.  We present a comprehensive $UBVRIJHK$-proper motion
catalogue for 2393 stars down to about $V=21.4$ mag observed in a $\sim 13'\times13'$
field centered on the cluster.
Fundamental parameters, such as core and cluster radius, reddening $E(B-V)$, age, 
distance modulus and mean proper motion were obtained using 
optical and near-IR photometry and proper motions. We analysed the cluster membership
using criteria based on distance from the cluster center, position in the CMD, and 
proper motions.  The membership probabilities of all stars in the field of
the cluster are presented.  We found 22 stars which are the most probable
cluster members. Our study indicates a distance of $2.43\pm0.23$\,kpc, a
diameter of $6.4\pm0.4$\,pc and an age of $224\pm25$\,Myr. The cluster is found 
to be heavily reddened with $E(B-V)=0.82\pm0.02$. The mean proper motion was estimated
to be $\mu_x = -3.29\pm0.35$\,mas\,yr$^{-1}$, $\mu_y = -1.24\pm0.28$\,mas\,yr$^{-1}$. Our 
analysis suggests that the cluster is slightly younger and more reddened 
than previously thought. It is important to note that because we limit  determinations 
to the most probable cluster members, the errors in the estimates of various 
cluster parameters has been considerably reduced.

The reddening law in the direction of the cluster was found to be normal at 
longer wavelength but anomalous at shorter wavelengths. In general, we found a 
slightly higher total-to-selective extinction $R_V=3.3$ towards NGC\,559. The
larger value of $R_V$ could be caused by a bigger than average grain size. 
Polarimetric data would be useful to ascertain the size and behaviour of the 
dust grains. From the combined optical and near-infrared  data, we obtained
a colour excesses of $E(V-K) = 2.14\pm0.02$, $E(J-K) = 0.37\pm0.01$, and 
$E(B-V)= 0.76\pm0.04$, in the direction of NGC\,559.

The MF for MS stars in the cluster is not uniform over the entire region
and found in the range $-1.64 \geq \Gamma \geq -2.14$ for the mass range
$1.0 \le M/M_\odot < 3.7$.  The MF slope of the core region is in agreement
with the Salpeter value, but it is found steeper in the corona and in the
cluster as a whole. This suggests mass segregation in MS stars due
to the dynamical evolution of the cluster. A deficiency of low-mass stars
as well as very massive stars was found in the core region of the cluster.
The age of the cluster was found to be much higher than the relaxation
time of 19.2\,Myr, which implies that the cluster is dynamically relaxed.
An improvement in the cluster parameters and knowledge of dynamical evolution
should allow a better understanding of star formation in NGC\,559.

In a forthcoming paper, we will report on stellar variability in NGC\,559
from 35 nights taken over 3 years during 2010 to 2012.
\section*{Acknowledgments}
The authors thank the anonymous referee for useful comments that improved the
scientific content of the paper. We acknowledge the suggestions given by Dr. Ramakant
Singh Yadav. This study has made use of data from the Two Micron All Sky Survey,
which is a joint project of the University of Massachusetts; the Infrared Processing
and Analysis Center/California Institute of Technology, funded by the National
Aeronautics and Space Administration and the National Science Foundation.

\label{lastpage}

\end{document}